\documentclass[conference, 10pt, final]{IEEEtran}
\usepackage[nolist]{acronym} 
\usepackage{tikz}
\usepackage{pgfplots}
\usepgfplotslibrary{groupplots}
\usepackage{graphicx}
\usepackage{subfloat}
\usepackage{caption}
\usepackage{subcaption}
\usepackage{float}
\pgfplotsset{compat=newest}
\usepackage{units}
\usepackage{amsmath, amsbsy, amssymb}
\usepackage{tikzscale}
\usetikzlibrary{arrows}
\usepackage{color}
\usepackage{epigraph}
\usepackage{circuitikz}
\usepackage{multirow}
\usepackage{booktabs}
\usepackage[plainruled]{algorithm2e}
\usepackage[group-separator={,}]{siunitx}
\definecolor{darkgreen}{rgb}{0.125,0.5,0.169}
\usetikzlibrary{shapes,arrows}
\usepackage{marvosym}
\usepackage{bbm}
\usepackage{mathtools}

\usepgfplotslibrary{fillbetween}
\usetikzlibrary{matrix}
\usetikzlibrary{dsp,chains}
\usetikzlibrary{shapes.geometric}

\tikzset{>=latex}

\newcommand{\Cc}{{\cal C}}

\newcommand{\Nc}{{\cal N}}

\begin{acronym}
 \acro{CSI}{channel state information}
 \acro{UE}{user equipment}
 \acro{UL}{uplink}
 \acro{BS}{basestation}
 \acro{TDD}{time division duplex}
 \acro{FDD}{frequency division duplex}
 \acro{ECC}{error-correcting code}
 \acro{MLD}{maximum likelihood decoding}
 \acro{HDD}{hard decision decoding}
 \acro{IF}{intermediate frequency}
 \acro{RF}{radio frequency}
 \acro{SDD}{soft decision decoding}
 \acro{NND}{neural network decoding}
 \acro{CNN}{convolutional neural network}
 \acro{ML}{maximum likelihood}
 \acro{GPU}{graphical processing unit}
 \acro{BP}{belief propagation}
 \acro{LTE}{Long Term Evolution}
 \acro{BER}{bit error rate}
 \acro{SNR}{signal-to-noise-ratio}
 \acro{ReLU}{rectified linear unit}
 \acro{BPSK}{binary phase shift keying}
 \acro{QPSK}{quadrature phase shift keying}
 \acro{AWGN}{additive white Gaussian noise}
 \acro{MSE}{mean squared error}
 \acro{LLR}{log-likelihood ratio}
 \acro{MAP}{maximum a posteriori}
 \acro{NVE}{normalized validation error}
 \acro{BCE}{binary cross-entropy}
 \acro{CE}{cross-entropy}
 \acro{BLER}{block error rate}
 \acro{SQR}{signal-to-quantisation-noise-ratio}
 \acro{MIMO}{multiple-input multiple-output}
 \acro{OFDM}{orthogonal frequency division multiplex}
 \acro{RF}{radio frequency}
 \acro{LOS}{line of sight}
 \acro{NLoS}{non-line of sight}
 \acro{NMSE}{normalized mean squared error}
 \acro{CFO}{carrier frequency offset}
 \acro{SFO}{sampling frequency offset}
 \acro{IPS}{indoor positioning system}
 \acro{TRIPS}{time-reversal IPS}
 \acro{RSSI}{received signal strength indicator}
 \acro{MIMO}{multiple-input multiple-output}
 \acro{ENoB}{effective number of bits}
 \acro{AGC}{automated gain control}
 \acro{ADC}{analog to digital converter}
 \acro{ADCs}{analog to digital converters}
 \acro{FB}{front bandpass}
 \acro{FPGA}{field programmable gate array}
 \acro{JSDM}{Joint Spatial Division and Multiplexing}
 \acro{NN}{neural network}
 \acro{IF}{intermediate frequency}
 \acro{LoS}{line-of-sight}
 \acro{NLoS}{non-line-of-sight}
 \acro{DSP}{digital signal processing}
 \acro{AFE}{analog front end}
 \acro{SQNR}{signal-to-quantisation-noise-ratio}
 \acro{SINR}{signal-to-interference-noise-ratio}
 \acro{ENoB}{effective number of bits}
 \acro{AGC}{automated gain control}
 \acro{PCB}{printed circuit board}
 \acro{EVM}{error vector mangnitude}
 \acro{CDF}{cumulative distribution function}
 \acro{MRC}{maximum ratio combining}
 \acro{MRP}{maximum ratio precoding}
 \acro{MRT}{maximum ratio transmission}
 \acro{DeepL}{deep-learning}
 \acro{DL}{deep learning}
 \acro{SISO}{single-input single-output}
 \acro{SGD}{stochastic gradient descent}
 \acro{CP}{cyclic prefix}
 \acro{MISO}{Multiple Input Single Output}
 \acro{LMMSE}{linear minimum mean square error}
 \acro{ZF}{zero forcing}
 \acro{USRP}{universal software radio peripheral}
 \acro{RNN}{recurrent neural network}
 \acro{GRU}{gated recurrent unit}
 \acro{LSTM}{long short-term memory}
 \acro{NTM}{neural turing machine}
 \acro{DNC}{differentiable neural computer}
 \acro{TCN}{temporal convolutional network}
 \acro{FCL}{fully connected layer}
 \acro{MANN}{memory augmented neural network}
 \acro{RNN}{recurrent neural network}
 \acro{DNN}{dense neural network}
 \acro{FIR}{finite impulse response}
 \acro{BPTT}{back-propagation through time}
 \acro{GAN}{generative adversarial network}
 \acro{ELU}{exponential linear unit}
 \acro{tanh}{hyperbolic tangent}
 \acro{BICM}{bit-interleaved coded modulation}
 \acro{OTA}{over-the-air}
 \acro{IM}{intensity modulation}
 \acro{DD}{direct detection}
 \acro{RL}{reinforcement learning}
 \acro{SDR}{software-defined radio}
 \acro{WGAN}{Wasserstein generative adversarial network}
 \acro{BMD}{bit-metric decoding}
 \acro{BMI}{bit-wise mutual information}
 \acro{LDPC}{low-density parity-check}
 \acro{IDD}{iterative demapping and decoding}
 \acro{JSD}{Jensen-Shannon divergence}
 \acro{MMSE}{minimum mean square error}
 \acro{FFT}{fast Fourier transform}
 \acro{IFFT}{inverse fast Fourier transform}
 \acro{QAM}{quadrature amplitude modulation}
 \acro{EMD}{earth mover's distance}
 \acro{TDL}{tapped delay line}
 \acro{KL}{Kullback-Leibler}
  \acro{IDFT}{inverse discrete Fourier transform}
 \acro{DFT}{discrete Fourier transform}
 \acro{PDP}{power delay profile}
 \acro{ISI}{intersymbol interference}
 \acro{LS}{least squares}
 \acro{V2X}{Vehicle-to-Everything}
 \acro{TDDL}{time-distributed dense layer}
 \acro{APP}{a posterior probability}
 \acro{WSSUS}{wide-sense stationary uncorrelated scattering}
\end{acronym}

\definecolor{mittelblau}{RGB}{0, 126, 198}
\definecolor{violettblau}{cmyk}{0.9, 0.6, 0, 0}
\definecolor{rot}{RGB}{238, 28 35}
\definecolor{apfelgruen}{RGB}{140, 198, 62}
\definecolor{gelb}{RGB}{1, 221, 0}
\definecolor{orange}{RGB}{244, 111, 33}
\definecolor{pink}{RGB}{237, 0, 140}
\definecolor{lila}{RGB}{128, 10, 145}
\definecolor{hellgrau}{RGB}{224, 224, 224}
\definecolor{mittelgrau}{RGB}{128, 128, 128}
\definecolor{dunkelgrau}{RGB}{80,80,80}
\definecolor{anthrazit}{RGB}{19, 31, 31}

\pgfplotscreateplotcyclelist{corporate colours markers}{%
rot, every mark/.append style={fill=.!80!rot},mark=*\\%
mittelblau, every mark/.append style={fill=.!80!mittelblau},mark=square*\\%
apfelgruen, every mark/.append style={fill=.!80!apfelgruen},mark=triangle*\\%
orange, mark=star\\%
pink, every mark/.append style={fill=.!80!pink},mark=diamond*\\%
violettblau, every mark/.append style={fill=.!80!violettblau},mark=otimes*\\%
lila, mark=|\\%
gelb, every mark/.append style={fill=.!80!gelb},mark=pentagon*\\%
hellgrau, mark=text,text mark=p\\%
anthrazit, mark=text,text mark=a\\%
}

\pgfplotscreateplotcyclelist{corporate colours markers double}{%
rot, every mark/.append style={fill=.!80!rot},mark=*\\%
rot, dashed, every mark/.append style={fill=.!80!rot,solid},mark=*\\%
mittelblau, every mark/.append style={fill=.!80!mittelblau},mark=square*\\%
mittelblau, dashed, every mark/.append style={fill=.!80!mittelblau,solid},mark=square*\\%
apfelgruen, every mark/.append style={fill=.!80!apfelgruen},mark=triangle*\\%
apfelgruen, dashed, every mark/.append style={fill=.!80!apfelgruen,solid},mark=triangle*\\%
orange, mark=star\\%
orange, dashed, mark=star\\%
pink, every mark/.append style={fill=.!80!pink},mark=diamond*\\%
pink, dashed, every mark/.append style={fill=.!80!pink,solid},mark=diamond*\\%
violettblau, every mark/.append style={fill=.!80!violettblau},mark=otimes*\\%
violettblau, dashed, every mark/.append style={fill=.!80!violettblau,solid},mark=otimes*\\%
lila, mark=|\\%
lila, dashed, mark=|\\%
gelb, every mark/.append style={fill=.!80!gelb},mark=pentagon*\\%
gelb, dashed, every mark/.append style={fill=.!80!gelb,solid},mark=pentagon*\\%
}

\IEEEoverridecommandlockouts

\usetikzlibrary{external}
\begin{document}

\title {Wiener Filter versus Recurrent Neural Network-based 2D-Channel Estimation for V2X Communications}

\author{\IEEEauthorblockN{Moritz Benedikt Fischer$^{1}$, Sebastian D\"orner$^{1}$, Sebastian Cammerer$^{1}$,\\Takayuki Shimizu$^{2}$, Bin Cheng$^{2}$, Hongsheng Lu$^{2}$, and Stephan ten Brink$^{1}$\\}

\IEEEauthorblockA{
$^{1}$ Institute of Telecommunications, Pfaffenwaldring 47, University of  Stuttgart, 70569 Stuttgart, Germany \\ \{fischer,doerner,cammerer,tenbrink\}@inue.uni-stuttgart.de\\
$^{2}$ InfoTech Lab, Toyota Motor North America \\ \{takayuki.shimizu,bin.cheng,hongsheng.lu\}@toyota.com\\
}

\thanks{This work has been supported by Toyota Motor North America and by the Federal Ministry of Education and Research of the Federal Republic of Germany through the FunKI project under grant 16KIS1187.}
}

\maketitle

\begin{abstract}
We compare the potential of \ac{NN}-based channel estimation with \emph{classical} \ac{LMMSE}-based estimators, also known as Wiener filtering.
For this, we propose a low-complexity \ac{RNN}-based estimator that allows channel equalization of a sequence of channel observations based on independent time- and frequency-domain \ac{LSTM} cells.
Motivated by \ac{V2X} applications, we simulate time- and frequency-selective channels with \ac{OFDM} and extend our channel models in such a way that a continuous degradation from \ac{LoS} to \ac{NLoS} conditions can be emulated.
It turns out that the \ac{NN}-based system cannot just compete with the \ac{LMMSE} equalizer, but it also can be trained w.r.t. resilience against system parameter mismatch.
We thereby showcase the conceptual simplicity of such a data-driven system design, as this not only enables more robustness against, e.g., \ac{SNR} or Doppler spread estimation mismatches, but also allows to use the \emph{same} equalizer over a wider range of input parameters without the need of re-building (or re-estimating) the filter coefficients.
Particular attention has been paid to ensure compatibility with the existing IEEE 802.11p piloting scheme for \ac{V2X} communications. %
Finally, feeding the payload data symbols as additional equalizer input unleashes further performance gains.
We show significant gains over the conventional \ac{LMMSE} equalization for highly dynamic channel conditions if such a \emph{data-augmented} equalization scheme is used.
\end{abstract}

\section{Introduction}

\begin{figure*}[]
 		\centering
 		\input{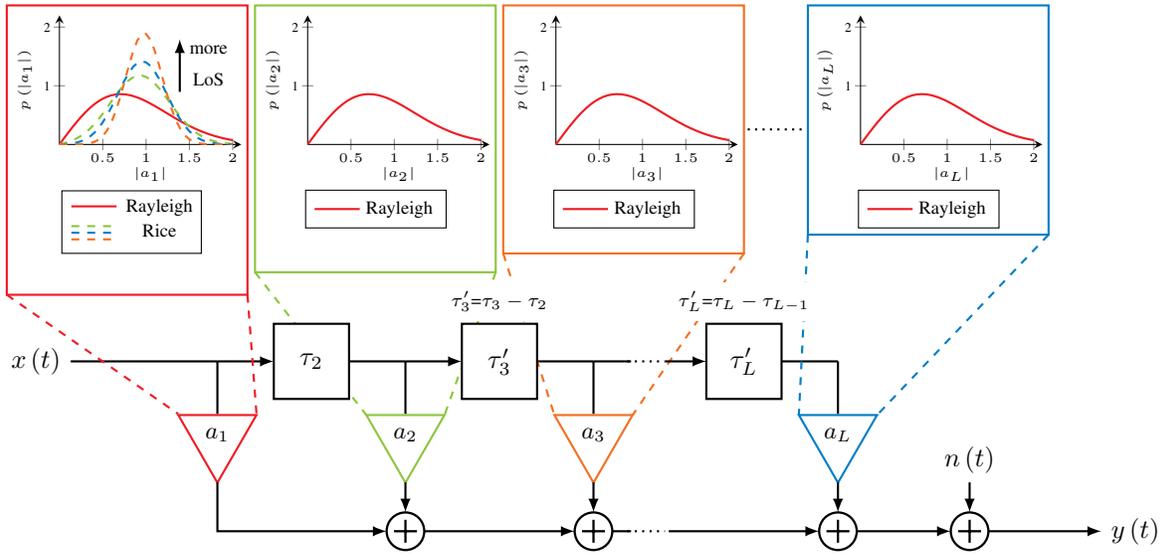}
		\caption{Tapped-delay line channel model with the delay of the first channel tap set to zero.}
		\label{fig:TDL_model}
\end{figure*}

In wireless communications, high-spectral efficient data transmission strongly depends on the availability of precise knowledge of the current \ac{CSI}. 
Thus, the \ac{CSI} needs to be estimated at the receiver -- either explicitly or implicitly. 
In particular, for channels that are time- and frequency-selective due to motion and multi-path propagation, respectively, channel estimation becomes a key challenge in today's \ac{OFDM}-based transceiver implementations.
Although channel estimation is a well-known field of \emph{classical} signal processing-based research, its performance is limited by the following challenges: (a) mismatch between the mathematical traceability of the actual physical channel, i.e., an inaccuracy of the underlying channel models, (b) computational complexity and also the induced latency of the algorithms, and (c) robustness w.r.t. the system parameter estimation such as \ac{SNR} or velocity.
All of this leads to a discrepancy between the theoretical optimal solution and the actually achieved performance of channel equalization when deployed in a real system.

Although the channel changes continuously, the amount of transmitted pilots used for channel estimation should be as low as possible to maximize the transmission rate. Obviously, the channel state in-between pilot positions must be interpolated.
Several of such pilot-aided channel estimation methods have evolved and essentially differ in the quality of this interpolation, but also in the required availability of assumptions about the channel, i.e., how is the channel correlated over time and frequency. A simple channel estimator is the \ac{LS} estimator with bilinear interpolation.
The 2D-\ac{LMMSE} estimator is a more advanced linear filter, which yields an optimal estimate of the channel under certain conditions \cite{nilsson1997analysis,hoeher1997two}; however, it is computationally complex and relies on the knowledge of the channel's statistical properties, which boils down to the task of estimating the channel covariance matrix.
To reduce the complexity, it can be replaced by two sequentially executed 1D-\ac{LMMSE} estimators \cite{dong2007linear}. 

Recently, deep learning for communications has attracted a lot of attention in academia and also in industry for virtually any possible application \cite{8054694,farsad2018neural,nachmani2016learning}.
This new paradigm of a data-centric system design allows to \emph{learn} equalizers that are perfectly aware of the channel statistics including all its (potential) impairments.
In \cite{neumann2018learning} an \ac{NN} structure for channel estimation is derived from the \ac{MMSE} filter. The main focus of \cite{luo2018channel} lies on the usage of additional meta data to obtain a more accurate channel estimate. Further, learning an efficient pilot arrangement together with a channel estimator for massive \ac{MIMO} has been done in \cite{mashhadi2020pruning}. A joint demapping and decoding scheme is presented in \cite{honkala2020deeprx} for a complete message frame. Further investigation on joint estimation and demapping as well as the transition towards end-to-end learning of the whole system including learning of the pilot arrangement and superimposed pilots are analyzed in \cite{aoudia2020end}. It has been observed in \cite{honkala2020deeprx} and \cite{aoudia2020end} that a \emph{data-aware} channel estimation can further improve the system's performance by analyzing the payload data besides considering only the specific pilot positions. Finally, the authors of \cite{li2018power} propose an \ac{NN}-based \ac{OFDM} receiver that implicitly estimates the required \ac{CSI}.

The main focus of our work is to showcase and analyze the \emph{universality} and \emph{flexibility} of \ac{NN}-based solutions.
For this, we seek to train our network over a wide range of possible input parameters. Motivated by the result in \cite{tandler2019recurrent}, which showed that \acp{RNN} are an attractive architecture for sequential data processing in communications, a novel \ac{RNN}-based channel estimator is presented in this work. 
The presented channel estimation scheme is evaluated over time- and frequency-selective channels and compared to conventional channel estimation techniques such as 2D-\ac{LMMSE} and 2x1D-\ac{LMMSE}. For this, we use the piloting scheme as given in the IEEE 802.11p standard \cite{ieee2010802}.
By investigating the influence of channel parameters such as velocity, noise and strength of a possible \ac{LoS} connection on the accuracy of the estimated channel, we study whether it is beneficial to use an \ac{RNN}-based channel estimator instead of conventional channel estimation schemes.
Besides this, we analyze to what extent gains are possible while relying on inaccurate knowledge of the statistical properties and parameters of the channel. 
In the last section, we show that a \emph{data-augmented} version of the same \ac{RNN} equalizer can further extract statistical information about the channel state by observing the received payload data symbols.

\section{System setup}
\subsection{Channel model}

We assume a time- and frequency-selective wireless channel described by its time-variant channel impulse response
	\begin{equation}
	\label{eq:tdl_channel}
		h\left(t, \tau\right) = \sum_{\ell=1}^{L}a_{\ell}\left(t\right)\delta\left(\tau-\tau_\ell\right)
	\end{equation}
where $L$ describes the number of channel taps, $a_\ell \left(t\right)$ denotes the complex-valued gain with an average power $p_\ell$ and $\tau_\ell$ represents the delay of the $\ell$-th channel tap. Each channel tap corresponds to one resolvable subpath of all paths in the multi-path environment, which is common for \ac{V2X} communications.
If transmitter and receiver are physically separated and there is no line-of-sight between them, the channel gain $a_\ell \left(t\right)$ is complex Gaussian distributed with zero-mean resulting in Rayleigh fading channel taps. This is a well-suited assumption if all received signal components are reflected or diffracted by objects of the environment. For further details we refer the reader to {\cite{Molisch, Stueber}. 
In case there is a strong \ac{LoS} component, it is more appropriate to model the first channel tap as Rician fading. The $K$-factor $K$ of the model is defined as the ratio between the power of the \ac{LoS} component and the power of all the remaining resolvable multi-path components. 
We follow the simulation framework as described in \cite{Xiao}.

Throughout this work, we assume \ac{OFDM} with \ac{CP} leading to the possibility of single-tap equalization. Let $X_{k,n}$ be the $k$-th complex symbol that should be transmitted over the $n$-th sub-carrier, where $k\in \mathbb{N}$ is the index of discrete time  and $n\in \lbrace0,..., N_{\mathrm{Sub}}-1 \rbrace$ is the sub-carrier over which the symbol should be transmitted. The symbol $X_{k,n}$ can either be a data carrying symbol or a pilot symbol that can be utilized for the channel estimation. The pilot arrangement for the IEEE 802.11p standard is schematically shown in Fig.~\ref{fig:80211p_pilot_arrangement}.
The mapping of the symbols $X_{k,n}$ onto the orthogonal sub-carriers can be expressed by an \ac{IDFT} and to eliminate \ac{ISI} a \ac{CP} is used. %
Let $\underline{X}_{k} = (X_{k,0}, X_{k,1}, ... , X_{k,N_{\mathrm{Sub}}-1})^T$ denote the complex-valued vector containing the symbols per sub-carrier of the $k$-th \ac{OFDM} symbol and $\underline{s}_{k} = (s_{k,0}, s_{k,1}, ... , s_{k,N_{\mathrm{Sub}}-1})^T$ is the complex vector containing $N_{\mathrm{Sub}}$ output samples of the \ac{IDFT} representing the $k$-th \ac{OFDM} symbol in the time domain.\footnote{For simplicity, we neglect the \ac{CP} in our description.}

	\begin{figure}[]
		\centering
 		\resizebox{!}{6.5cm}{\input{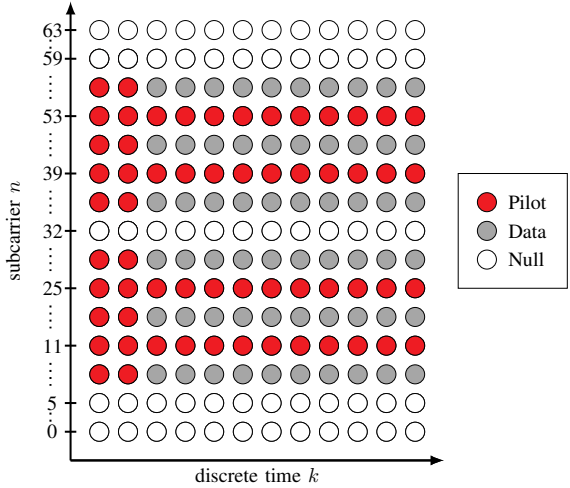}}
		\caption{IEEE 802.11p pilot arrangement.}
		\label{fig:80211p_pilot_arrangement}
	\end{figure}

If the maximal delay of the channel is smaller than the duration of the \ac{CP}, the received symbols $\underline{r}_{k}$ in the time domain after the removal of the \ac{CP} are given by
	\begin{equation}
		\underline{r}_{k} = \underline{h}_{k} \circledast \underline{s}_{k} + \underline{n}_k
	\label{eq:channel_circ_conv}
	\end{equation} 
where $\underline{h}_k$ is the sampled version of the channel impulse response $h\left(t, \tau \right)$, $\underline{n}_k$ denotes an \ac{AWGN} term and $\circledast$ denotes the circular convolution. The received symbols in the frequency domain are defined as	
	\begin{equation}
		\underline{Y}_k {=} \underline{H}_{k} \circ \underline{X}_{k}+ \underline{N}_k.
	\label{eq:channel_DFT}
	\end{equation}%
where $\underline{H}_{k}$ is the sampled channel transfer function, i.e., the \ac{DFT} of the sampled channel impulse response $\underline{h}_k$.
Furthermore, $\circ$ denotes the Hadamard product, i.e., the element-wise multiplication.
In the following we assume that the individual entries of $\underline{N}_k$ are independent and identically complex Gaussian distributed with zero-mean and a variance of $\sigma^2$. Hence, $\underline{N}_k$ is  	
	\begin{equation}
	\label{eq:noise_vector}
		\underline{N}_k \sim \mathcal{C}\mathcal{N}(\underline{0},\sigma^2\mathbf{I}_{N_{\mathrm{Sub}}})
	\end{equation} 
distributed, where $\mathbf{I}_{N_{\mathrm{Sub}}}$ is the identity matrix of size $N_{\mathrm{Sub}} \times N_{\mathrm{Sub}}$. 	

At the receiver the symbols $\underline{Y}_k$ are equalized by scaling such that %
	\begin{equation}
	\label{eq:OFDM_EQ}
	\hat{Y}_{k,n} = Y_{k,n} \frac{\hat{H}_{k,n}^*}{|\hat{H}_{k,n}|^2}.
	\end{equation}
For the consideration of \ac{OFDM} frames of $n_\mathrm{T}$ \ac{OFDM} symbols (\ref{eq:channel_DFT}) can be extended to	
	\begin{equation}
	\label{eq:channel_OFDM}
		\mathbf{Y} = \mathbf{H}\circ\mathbf{X} + \mathbf{N}
	\end{equation}	
where $\mathbf{Y}\in\Cc^{n_\mathrm{T}\times n_\mathrm{F}}$ is the received symbol matrix, $\mathbf{H}\in\Cc^{n_\mathrm{T}\times n_\mathrm{F}}$ is the channel matrix, 
$\mathbf{X}\in\Cc^{n_\mathrm{T}\times n_\mathrm{F}}$ is the transmitted symbol matrix and $\mathbf{N}\in\Cc^{n_\mathrm{T}\times n_\mathrm{F}}$ is the \ac{AWGN} noise matrix.

	\begin{figure}[]
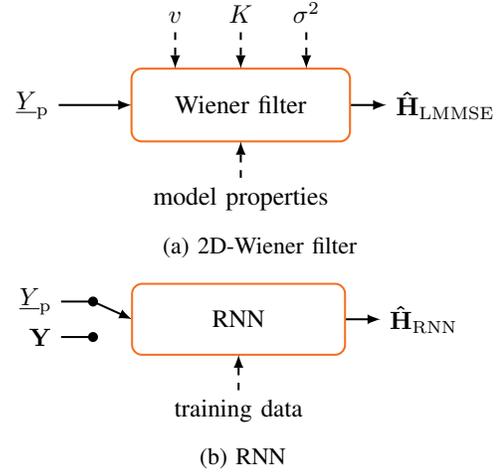

	\begin{subfigure}{0.5\textwidth}
 		\centering
 		\resizebox{!}{3cm}{\input{fig/Wiener_Filter.tex}}
		\caption{2D-Wiener filter}
		\label{fig:Wiener filter}
	\end{subfigure}
	\begin{subfigure}{.475\textwidth}
		\vspace{0.25cm}
		\centering
 		\resizebox{!}{2cm}{\input{fig/RNN_Inputs.tex}}
		\caption{\ac{RNN}}
		\label{fig:RNN}
	\end{subfigure}	
	\caption{Block diagram of the 2D-Wiener filter and the \ac{RNN}-based channel estimator.}
	\end{figure}
	
	\begin{figure*}[]
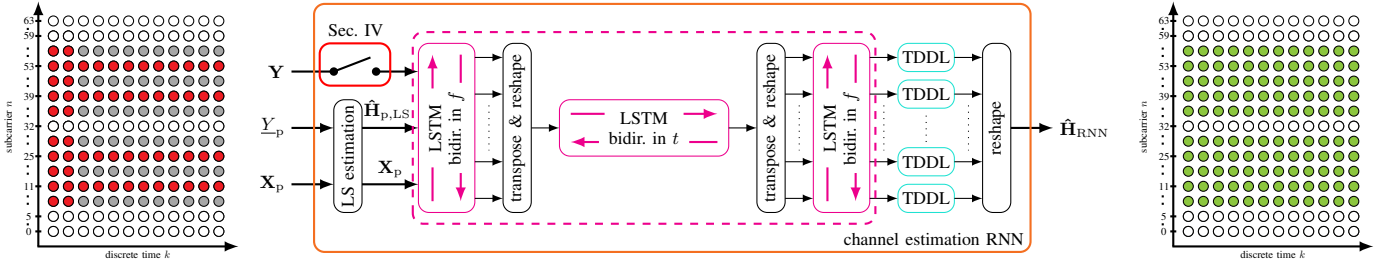

	\begin{center}
	\captionsetup[subfigure]{labelformat=empty}
\hspace{-1.cm}
		\begin{subfigure}{.17\textwidth}
 			\centering
 			\input{fig/pilot_arrangement_802_11p-subfigure.tex}
		\end{subfigure}
\hspace{-0.5cm}
		\begin{subfigure}{0.69\textwidth}
			\centering
			\input{fig/Fig_TF-RNN_channel_estimator_switch2.tex}
 			\vspace{0.5cm}
		\end{subfigure}
\hspace{-0.5cm}
	\begin{subfigure}{.14\textwidth}
 			\centering
 			\input{fig/pilot_arrangement_802_11p-estimate-subfigure.tex}
		\end{subfigure}	
	\end{center}	
 		\vspace*{-0.5cm}
	\caption{Block diagram of the RNN-based channel estimator. The additional input $\mathbf{Y}$ is only used in Sec.~\ref{sec:aug} for \emph{data-augmented} channel estimation.}
	\label{fig:RNN_block_diagram}
	\end{figure*}

\subsection{Wiener filter baseline}
The Wiener filter as shown in Fig.~\ref{fig:Wiener filter}, also denoted as \ac{LMMSE} estimator, is a linear filter that minimizes the \ac{MSE} between the true channel and the estimated channel. Therefore, it utilizes received pilot symbols, an estimated \ac{SNR} and knowledge of the statistical properties of the channel. Precisely, it needs to know the channel's autocorrelation function and the noise variance. 
For the derivation of the 2D-Wiener filter (\ref{eq:channel_OFDM}) is rewritten into 
	\begin{align}
 	\label{eq:channel_eq}
 	\underline{Y} &= \mathbf{X} \underline{H} + \underline{N}\\
 				&= \operatorname{diag}\left(\operatorname{vec}\left(\mathbf{X}\right)\right) \operatorname{vec}\left(\mathbf{H}\right)+ \operatorname{vec}\left(\mathbf{N}\right)
 	\end{align}
where $\mathbf{X}$ is a diagonal matrix with all transmitted symbols on the main diagonal and $\underline{Y}$, $\underline{H}$ and $\underline{N}$ are the vectorized versions of the received symbols, channel matrix and the \ac{AWGN} matrix, respectively. In the same way the received pilot symbols are defined as  	
	\begin{align}
 	\label{eq:channel_eq}
 	\underline{Y}_\mathrm{p} &= \mathbf{X}_\mathrm{p} \underline{H}_\mathrm{p} + \underline{N}_\mathrm{p}\\
 				&= \operatorname{diag}\left(\Pi \operatorname{vec}\left(\mathbf{X}\right)\right) \Pi \operatorname{vec}\left(\mathbf{H}\right)+ \Pi \operatorname{vec}\left(\mathbf{N}\right)
 	\end{align}
where $\Pi\in\Nc^{p \times n_\mathrm{T}n_\mathrm{F} }$ is a pilot selection matrix and $p$ is the number of pilots distributed over the \ac{OFDM} frame. The pilot selection matrix contains a single one per row to select the pilot position and all the other elements of a row are set to zero.   			

Following \cite{nilsson1997analysis} and \cite{hoeher1997two} the channel estimate of the 2D-Wiener filter is given by	
	\begin{equation}
	\label{eq:H_LMMSE}
		\hat{\underline{H}}_\mathrm{LMMSE} = \mathbf{W} \mathbf{X}_{\mathrm{p}}^{-1}\underline{Y}_{\mathrm{p}}
	\end{equation}
with the definition of the filter matrix $\mathbf{W}$	
	\begin{equation}
		\label{eq:2D-LMMSE-filter-matrix}
		\mathbf{W} = \mathbf{R}_{\underline{H}\underline{H}} \Pi^T \left(\Pi\left(\mathbf{R}_{\underline{H}\underline{H}}+ \sigma^2 \mathbf{I}_{n_\mathrm{T}n_\mathrm{F}}\right)\Pi^T\right)^{-1}.
	\end{equation}
Note that the multiplication of $\mathbf{X}_{\mathrm{p}}^{-1}$ with $\underline{Y}_{\mathrm{p}}$ is the \ac{LS} estimate of the channel at pilot positions.  
$\mathbf{R}_{\underline{H}\underline{H}}$ denotes the autocorrelation matrix of the channel, which we assume to be a separable composition of the temporal and the spectral autocorrelation of the channel and can be expressed as the Kronecker product \cite{Simko_Kronecker}  	
	\begin{equation}
	\label{eq:ACF}
		\mathbf{R}_{\underline{H}\underline{H}} = \mathbf{R}_\mathrm{T} \otimes \mathbf{R}_\mathrm{F}.
	\end{equation}
This is a valid assumption for \ac{WSSUS} channel models. In general, the autocorrelation of unknown channels has to be determined empirically.

The temporal autocorrelation matrix $\mathbf{R}_\mathrm{T}$ and the spectral autocorrelation matrix $\mathbf{R}_\mathrm{F}$ are both Toeplitz matrices with their elements given by the temporal autocorrelation function	
	\begin{equation}
	\label{eq:ACF_Time}
		r_\mathrm{T} \left(kT_\mathrm{S}\right) = \operatorname{J_0}\left(2 \pi f_\mathrm{D, max} T_\mathrm{S} k\right)
	\end{equation}
and the spectral autocorrelation function	
	\begin{equation}
	\label{eq:ACF_Freq}
		r_\mathrm{F} \left(m\Delta f\right) = \sum_{l=0}^{L-1} p_l e^{j2 \pi \tau_l \Delta f m},	
	\end{equation}
respectively. Hereby, $T_\mathrm{S}$ is the duration of an \ac{OFDM} symbol with \ac{CP}, $f_\mathrm{D, max}=\nicefrac{v}{c_0}f_\mathrm{c}$ denotes the maximal Doppler frequency of the channel, $v$ denotes the velocity, $c_0$ is the wave propagation speed, i.e., speed of light, $f_\mathrm{c}$ is the carrier-frequency, $\Delta f$ is the sub-carrier spacing and $\operatorname{J_0}\left(.\right)$ is the zero-th order Bessel function of the first kind.  	

To reduce the computational complexity of the 2D-Wiener filter it can be replaced by a 2x1D-Wiener filter, considering that the two-dimensional channel estimation problem is separated into two one-dimensional tasks that are solved subsequently. The first Wiener filter interpolates the channel over the time axis, and the second Wiener filter estimates the channel over the frequency axis. For further details we refer the reader to \cite{dong2007linear}, \cite{nilsson1997analysis} and \cite{hoeher1997two}.

As can be seen from (\ref{eq:2D-LMMSE-filter-matrix}), (\ref{eq:ACF_Time}) and (\ref{eq:ACF_Freq}) and as illustrated in Fig. \ref{fig:Wiener filter}, the filter matrices of the 2D-Wiener filter and the 2x1D-Wiener filter are dependent on several parameters of the channel, including velocity, noise power and $K$-factor. This additional knowledge of the channel parameters has to be provided to the Wiener filters by external parameter estimators. Furthermore, every time such a parameter changes, a recalculation of the filter matrices is required.

\section{RNN-based channel estimation}
\subsection{Neural network structure and training}
The structure of the proposed \ac{RNN}-based channel estimator is illustrated in Fig. \ref{fig:RNN_block_diagram} and mainly consists of 3 bidirectional recurrent \ac{LSTM} cells, which consecutively handle the input data in frequency-, then time- and then frequency-dimension again. This allows to keep the computational complexity feasible.
We use recurrent \acp{NN} to exploit temporal and frequency correlations of the input data, but instead of simple recurrent neurons, we use \ac{LSTM} units to circumvent the inherent vanishing gradient problem \cite{Hochreiter}.

The \ac{RNN} only takes the received pilot symbols $\underline{Y}_\mathrm{p}$, and the actually sent (known) pilot symbols $\mathbf{X}_\mathrm{p}$ as inputs for the two-dimensional channel estimation task.
Only in Sec.~\ref{sec:aug}, we additionally provide all the received symbols $\mathbf{Y}$ (including $\underline{Y}_\mathrm{p}$) to the \ac{RNN} for \emph{data-augmented} equalization. This is indicated by the red box in Fig.~\ref{fig:RNN_block_diagram}.

There are no explicit assumptions on channel statistics and noise variance fed to the \ac{RNN}, but all of this information is implicitly provided through the data-driven training approach (cf. Fig. \ref{fig:RNN}).
This is also the reason why we additionally feed $\mathbf{Y}$ (which includes random noisy payload data symbols) to the \ac{RNN} (only in Sec.~\ref{sec:aug}), as this allows the \ac{NN} to better \emph{learn} and estimate these channel statistics by itself.
The first preprocessing step of the \ac{RNN}-based estimator is the calculation of the \ac{LS} estimate at all pilot positions $\underline{\hat{H}}_\mathrm{p,LS} = \mathbf{X}_{\mathrm{p}}^{-1} \underline{Y}_\mathrm{p}$. Thereafter, the resulting vector is reshaped into a matrix $\mathbf{\hat{H}}_\mathrm{p,LS}$ of target shape $\Cc^{n_\mathrm{T} \times n_\mathrm{F}}$ and zeros are inserted at positions where data is transmitted.
Similar to \cite{honkala2020deeprx}, all inputs of shape $\Cc^{n_\mathrm{T} \times n_\mathrm{F}}$, including the pilot mask and positions in time and frequency dimension, are then stacked to one large input tensor.

After a complex- to real-valued conversion, the input data is processed by a bidirectional \ac{LSTM} cell in frequency direction (i.e., \emph{vertically} in the resource grid in Fig.~\ref{fig:80211p_pilot_arrangement}).
This means that the data (in frequency domain) is used as the \ac{LSTM} cell's time dimension. %
The output of the initial \ac{LSTM} cell is thus an estimation for each symbol, only dependent on input data for the specific symbol and incoming cell states from decisions on previous and subsequent sub-carrier symbols.
After the initial \ac{LSTM} cell, the outputs are reshaped and time and frequency dimensions are transposed, so that the second \ac{LSTM} cell can now operate in actual time domain (i.e., \emph{horizontally} in the resource grid in Fig.~\ref{fig:80211p_pilot_arrangement}).
This second \ac{LSTM} cell thereby has no direct connection in the frequency domain as its states only traverse trough the time domain.
Finally, the second cell's output is again reshaped and transposed so that a third \ac{LSTM} cell can operate on frequency domain again.

The third \ac{LSTM} cell's output is then further processed by two \acp{TDDL}, i.e., dense layers that are applied to each time-frequency step separately with the same shared weights.
The purpose of the first \ac{TDDL} layer, which is \ac{ReLU} activated, is to combine the bidirectional \ac{LSTM} cell's forward and backward output.
The second \ac{TDDL} layer consists of two linearly activated neurons which finally provide an estimation on the real and imaginary part of $H_{k,n}$.
We chose this 3x1D \ac{RNN} structure to minimize the \ac{NN}'s complexity for a single time-frequency step as much as possible with only recurrent state information traversing in time or frequency dimension.
This approach renders this architecture highly scalable for all kinds of input data dimensions.
Throughout this work, we use the proposed structure with 64 \ac{LSTM} units per cell and 8 neurons in the first \ac{TDDL} layer. %

During training of the \ac{RNN}-based channel estimator, \ac{SGD} and \ac{BPTT} are applied.
Since the channel estimation problem can be classified as a regression task, we use the \ac{MSE} loss; further training parameters are summarized in Tab.~\ref{tab:training_parameters}.
Note that, to obtain a universal channel estimator, the variety in the training data has to be high.
For this, the \ac{RNN} is trained within a given interval of uniformly distributed velocities and \ac{SNR} values.
In contrast to that, the $K$-factors used during training are not uniformly distributed, but instead, training data is divided into two halves.
One half of each training batch is created with \ac{NLoS} conditions, i.e., $K=0$ and the other half is created with uniformly distributed $K$-factors with $K \in (0,5)$.

	\begin{table}[]
		\centering
		\caption{Training parameters}
		\begin{tabular}{r|c} 
		\toprule
		Parameter & value\\ 
		\midrule 
		Epochs & 100\\
		Iterations per epoch & 1000 \\ 
		Batch size  & 200\\ 
		Learning rate & 0.001\\ 
		Velocities & 0$\unit{\frac{km}{h}}$ - 300$\unit{\frac{km}{h}}$\\
		\ac{SNR} & 5\unit{dB} - 30\unit{dB}\\
		$K$-factors & 0 - 5\\
		Fract. of pure \ac{NLoS} scenarios in train. set & 50\%\\
		\bottomrule
		\end{tabular}
		\label{tab:training_parameters}
	\end{table}
	
		\begin{table}[]
		\centering
		\caption{Simulation parameters}
		\begin{tabular}{r|c} 
		\toprule
		Parameter & value\\ 
		\midrule 
		Carrier frequency $f_\mathrm{c}$&  5.9 \unit{GHz} \\
		Signal bandwidth $B$& 10 \unit{MHz} \\ 
		\ac{OFDM} symbol duration with \ac{CP} $T_\mathrm{S}$& 8 $\unit{\mu s}$\\ 
		Cyclic prefix duration $T_\mathrm{CP}$& 1.6 $\unit{\mu s}$\\ 
		Sub-carrier spacing $\Delta f$& 156.25 $\unit{kHz}$\\
		Number of sub-carriers $N_\mathrm{Sub}$& 64\\
		Number of used sub-carriers & 52\\
		Pilot arrangement& IEEE 802.11p\\
		Modulation&  QPSK or 16-QAM\\
		Channel length $L$ & 6 taps\\
		Maximum delay $\tau_\mathrm{max}$&  500 \unit{ns}\\
		\Ac{PDP} &  equally weighted\\
		Fading statistics & Rayleigh or Rician\\
		\bottomrule
		\end{tabular}
		\label{tab:simulation_parameter}
	\end{table}
	
\subsection{Simulation results}

	\begin{figure}
 		\centering
 		\input{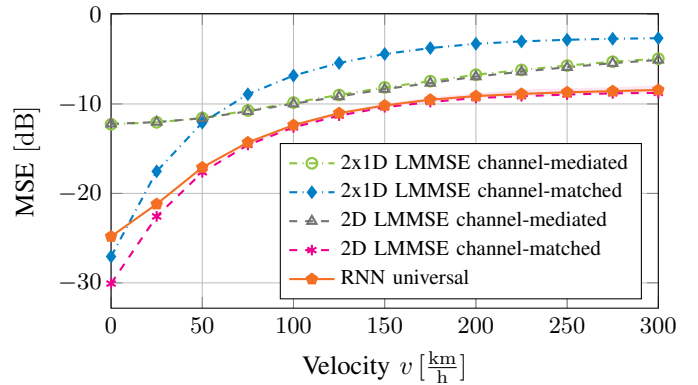}
 		\vspace*{-0.5cm}
		\caption{MSE comparison for different velocities at an \ac{SNR} of 15 \unit{dB} and a $K$-factor of $K=0$. The influence of up to $\pm 30$\% velocity estimation mismatch on the 2D-\ac{LMMSE} is visualized by the shaded red area around the \emph{LMMSE channel-matched curve}.}%
		\label {fig:V_MSE}
	\end{figure}

We consider the \ac{V2X} wireless channel with a carrier frequency of 5.9 \unit{GHz}. The parameters assumed for the simulations are guided by the IEEE 802.11p standard \cite{ieee2010802} and are shown in detail in Tab.~\ref{tab:simulation_parameter}. Thereby, a pilot arrangement as depicted in Fig.~\ref{fig:80211p_pilot_arrangement} is used.

In Fig.~\ref{fig:V_MSE} to Fig.~\ref{fig:Mismatch_MSE} the novel \ac{RNN}-based channel estimator is compared with the 2D-Wiener filter and the 2x1D-Wiener filter (with reduced computational complexity) for varying velocities, \ac{SNR} values and $K$-factors. The Wiener filters are recalculated for every evaluation point whereas the \ac{RNN}-based channel estimator does not change for varying channel parameters.
Besides the Wiener filters based on the accurate knowledge of the statistical properties and parameters of the channel (\emph{channel-matched} \ac{LMMSE}), more practicable estimated versions of them are also included (\emph{channel-mediated} \ac{LMMSE}) that are valid over a wider range of input parameters. Their filter matrices are estimated over $10^6 $ channel realizations for uniformly distributed velocities between $0\unit{\nicefrac{km}{h}}$ and $300\unit{\nicefrac{km}{h}}$, SNR values between $5\unit{dB}$ and $30\unit{dB}$ and $K$-factors between $K=0$ and $K=5$. They are assumed to be applicable for a range of channel parameters and, therefore, a recalculation of the Wiener filters at runtime is not needed anymore.
To summarize, the following filters are shown:
\begin{itemize}
\item \textbf{Channel-\emph{matched} \ac{LMMSE}} with genie-aided (perfect) parameter knowledge at the receiver. We show the 2D-\ac{LMMSE} version and its 2x1D simplification for scenarios where it provides further insights.
\item \textbf{Channel-\emph{mediated} \ac{LMMSE}} with estimated (averaged over many channel realizations) parameters, in an attempt to obtain an \emph{universal} \ac{LMMSE} estimator applicable over a wide range of parameters, serving as a best possible conventional competitor to the (later) universal \ac{RNN} approach.
\item \textbf{\ac{RNN} \emph{universal}} trained over a wide range of input parameters that does \emph{not} require any other explicit parameter knowledge during inference.
\end{itemize}

The results reveal that the \ac{RNN}-based channel estimator is universally applicable over a wide range of parameters and can adapt to varying and also difficult channel conditions, e.g., high user velocities as shown in Fig.~\ref{fig:V_MSE}. 
It can be seen that a high velocity renders the channel estimation into a difficult task and, thus, a general degradation of all estimators can be observed due to the high time-selectivity of the channel.
Somewhat to our surprise, the \ac{RNN}-based channel estimator can compete with the channel-matched \ac{LMMSE} equalizer without being fed with a velocity estimation. 
This can be intuitively explained by the practice of training over a wide range of different channel parameters and thereby forcing the \ac{RNN} to inherently \emph{learn} to handle these effects solely based on input received channel observations.
Note that the visible degradation of the 2x1D \ac{LMMSE} performance is mainly caused by this standard's pilot positioning scheme.%

	\begin{figure}
 		\centering
 		\input{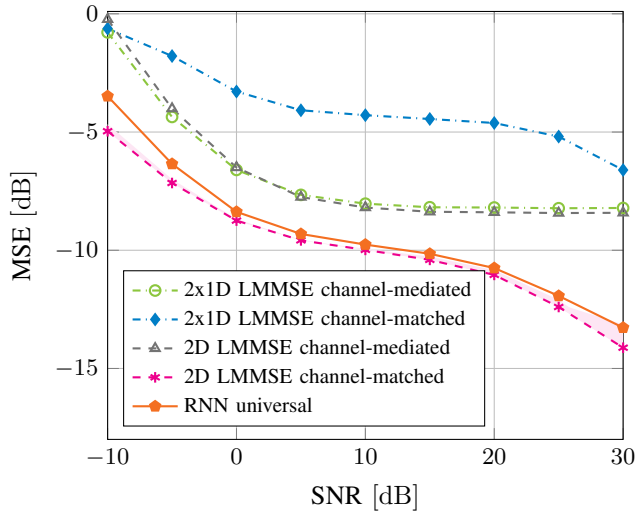}
 		\vspace*{-0.1cm}
		\caption{MSE comparison for different \acp{SNR}. The influence of up to $\pm 3$\unit{dB} \ac{SNR} estimation mismatch on the 2D-\ac{LMMSE} is visualized by the shaded red area around the \emph{LMMSE channel-matched curve}.
		}
		\label{fig:SNR_MSE}
	\end{figure}

Fig.~\ref{fig:SNR_MSE} analyzes the impact of the channel \ac{SNR} and the red region around the \ac{LMMSE} equalizer shows the influence of an $\pm 3\unit{dB}$ SNR parameter estimation mismatch. We want to emphasize, that the \ac{RNN}-based channel estimator has an inherent advantage compared to the Wiener filter, as it only takes received channel observations as input and no further parameter estimation is required. A recalculation during runtime, as well as a separate parameter estimation block, is not needed.
Therefore, this reduces its design complexity compared to a 2D-Wiener filter (assuming a continuous adaption of the filter coefficients to newly estimated parameters).
Its universality and the absence of a separate parameter estimation lead to the conjecture, that the \ac{RNN} implicitly learns to estimate these parameters.
This is also supported by the observation that it outperforms both channel-mediated methods, which also only use the received pilot symbols without additional knowledge of the exact parameters.
In cases where the 2D-Wiener filter is known to be optimal, the \ac{RNN} can compete with it and is, in the most relevant regions, on par with the 2D-Wiener filter in terms of \ac{MSE} performance. %

	\begin{figure}
 		\centering
 		\input{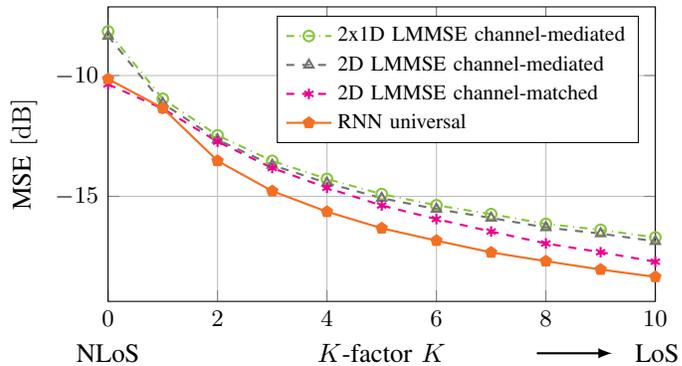}
 		\vspace*{-0.5cm}
		\caption{MSE comparison for different $K$-factors at a velocity of $150\unit{\nicefrac{km}{h}}$ and an \ac{SNR} of 15 \unit{dB}.}
		\label{fig:K_MSE}
	\end{figure}

In Fig.~\ref{fig:K_MSE} we now vary the channel properties by increasing the $K$-factor of our model. It is instructive to realize that this implies that the \ac{LoS} components increasingly dominate the channel's behavior.
Note that this does not necessarily render the problem more difficult (often \ac{LoS} conditions are much easier to handle; thus, the overall performance also improves for all the estimators with increasing $K$), but is simply unexpected by the equalizer and, thus, difficult to handle due to the model mismatch. Or in other words, we have designed the equalizer for the wrong channel conditions, which is beneficial for the \ac{RNN} that only extracts the model from training data itself.
We want to emphasize that this experiment operates the equalizers under mismatched conditions on purpose, i.e., we want to showcase the universality of the \ac{RNN}-based system.
Furthermore, in the case of evaluating the performance for varying $K$-factors and $K$-factors larger than zero, as shown in Fig.~\ref{fig:K_MSE}, the 2D-Wiener filter is not optimal anymore.
Although, it is still the best conventional channel estimation method for this task, the \ac{RNN} is able to outperform the 2D-Wiener filter for the given mismatch.
However, the importance of this experiment is not necessarily found in the pure performance gain, but also in the simplicity of the system design. 

	\begin{figure}
 		\centering
 		\input{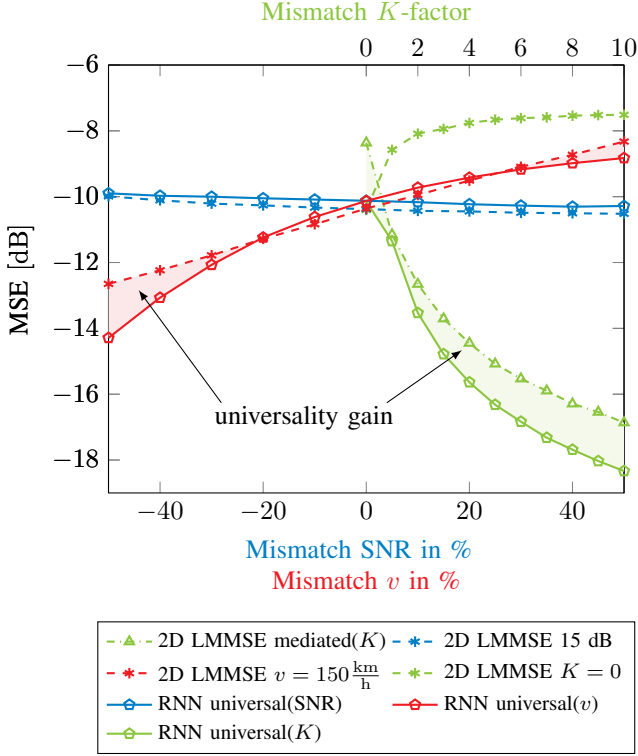}
		\caption{MSE comparison for the mismatch of the velocity, for the mismatch of the \ac{SNR} and for the mismatch of the $K$-factor (based on a velocity of $150\unit{\nicefrac{km}{h}}$, an \ac{SNR} of 15 \unit{dB} and a $K$-factor of 0).}
		\label{fig:Mismatch_MSE}
	\end{figure}	

These gains in universality are also visible in Fig.~\ref{fig:Mismatch_MSE}, where the \ac{RNN}'s \ac{MSE} performance is compared to a mismatched \ac{LMMSE} estimator, which assumes $K=0$, $v=150\unit{\nicefrac{km}{h}}$ and an \ac{SNR} of $15\unit{dB}$. The shaded areas visualize regions for which the \ac{RNN} provides a significantly better performance than the best \ac{LMMSE} version. For readability, other estimators are not shown here.
It can be clearly seen that the \ac{LMMSE} equalizer performs slightly better if all parameters are well known, however, when introducing a mismatch of \ac{SNR}, velocity or $K$-factor, the \ac{RNN} clearly outperforms the \emph{classical} approaches.

	\begin{figure}
 		\centering
 		\input{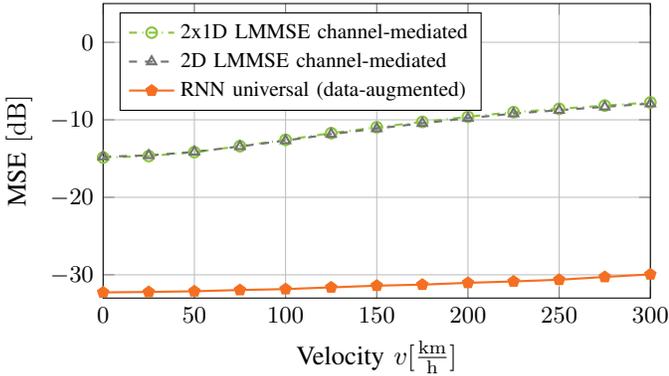}
 		\vspace*{-0.5cm}
		\caption{\Ac{MSE} comparison 
		for different velocities at randomly chosen $K$-Factors between 0 and 5 (at least 20\% $K=0$) and randomly chosen \acp{SNR} in the range between $5 \unit{dB}$ and $30\unit{dB}$. The \ac{RNN}-based equalizer now uses $\mathbf{Y}$ as additional input, i.e., \emph{data-augmented} equalization is used.}
		\label{fig:V_MSE_3x1D}
	\end{figure}

	\begin{figure}
 		\centering
 		\input{fig/SNR_BER-80211p.tex}
 		\vspace*{-0.5cm}
		\caption{\Ac{BER} vs. \ac{SNR} for \ac{QAM} constellations of $m=2$ and $m=4$ bit per complex symbol after soft-demapping and 40 iterations of \ac{BP} decoding.}
		\label{fig:SNR_BER}
	\end{figure}

\section{Data Augmented Equalization}
\label{sec:aug}

We now extend the \ac{RNN} equalizer by additionally providing the data symbols (without involving any decision feedback) to the equalizer similar to \cite{honkala2020deeprx,aoudia2020end} and, thereby, realize a \emph{data-aware} equalization.
This means we simply enable the additional input of $\mathbf{Y}$ in Fig.~\ref{fig:RNN_block_diagram} (red box) and re-train the system. Note that we only re-use existing data that must be available at the receiver anyway but remains unused in the \emph{classical} linear equalization schemes.\footnote{Note that we only provide the raw input $\mathbf{Y}$ without involving any further decisions as it is common for classical \emph{decision feedback} and \emph{iterative channel estimation and decoding} algorithms. However, the scheme could be extended to also account for decision feedback from the forward error correction as done in \cite{cammerer2020trainable}.}
The observed gains (shown in Fig.~\ref{fig:V_MSE_3x1D}-~\ref{fig:SNR_BER} and discussed in the following) can intuitively be explained by the fact that the data symbols -- although unknown at the receiver -- can only occur at specific positions of the regular \ac{QAM} constellation grid.
This information is typically not used, however, the conceptual elegance of \acp{NN} easily allows to extract further statistical information about the current channel state even with data suffering from high uncertainty (unknown data bits and additional channel noise).
So far the gains of the \ac{RNN}-based equalizer have been limited and could be found more in terms of \emph{universality} but not necessarily in terms of \ac{MSE} performance gains when compared to the \ac{LMMSE} equalizer. However, this experiment underlines the conceptual simplicity of such a \emph{data-driven} system design w.r.t. re-using additional data sources. We simply re-train the system with the additional data input and do not need any reconsideration of the system setup itself.
We empirically observe that this data augmentation provides significant additional gains in highly dynamic situations such as high velocity scenarios which would otherwise require more pilots.

Fig.~\ref{fig:V_MSE_3x1D} shows the \ac{MSE} performance at varying velocities in a scenario with $K$-factors randomly chosen between 0 to 5 with at least 20\% of \ac{NLoS} conditions ($K=0$) and \ac{SNR} randomly chosen between $5 \unit{dB}$ and 30$\unit{dB}$. Due to all channel parameters being randomly chosen, Fig.~\ref{fig:V_MSE_3x1D} only shows the channel-mediated estimators, with optimized filter coefficients for given parameter ranges, as conventional baseline.
Clearly, the additional input of the received data symbols $\mathbf{Y}$ helps the \ac{RNN} to track the variations of the channel and it can outperform both comparable channel-mediated 2x1D and 2D LMMSE baseline systems over the whole range of investigated velocities.%

To further highlight these advantages in terms of resilience and a more accurate channel estimation, we now examine whether the observed gains for \ac{MSE} on channel estimation can be transferred to the more tangible metric of final \ac{BER} performance.
Therefore, we use $\hat{\mathbf{H}}$ to employ \ac{APP} soft-demapping and add an outer %
irregular \ac{LDPC} code of rate $r = \nicefrac{1}{2}$, length $n = 1296$ bit and 40 iterations of \ac{BP} decoding.
We also evaluate the \ac{BER} performance in the more realistic channel scenario with parameters $K$ randomly chosen from 0 to 5 with at least 20\% of \ac{NLoS} ($K=0$) realizations and velocity randomly chosen between 0 and 300$\unit{\nicefrac{km}{h}}$ at varying \ac{SNR}. %
Fig.~\ref{fig:SNR_BER} shows the \ac{BER} performance vs. $\nicefrac{E_b}{N_0}$ for randomly generated bit, modulated with \ac{QPSK} ($m=2$ bit per complex symbol) and 16-\ac{QAM} ($m=4$) constellations.
As can be seen, the \ac{RNN}'s generally improved estimation $\mathbf{\hat{H}}_\text{RNN}$ in terms of \ac{MSE} also leads to improved final \ac{BER} performance.
Thereby it is shown, that soft-values which are demapped using $\mathbf{\hat{H}}_\text{RNN}$ provide more information to the \ac{BP} decoder at a given \ac{SNR} than soft-values which are demapped using the conventional \ac{LMMSE} estimator.
Finally, we can observe gains in this specific scenario with varying parameters of up to 1\unit{dB} for $m=2$ and even higher gains of up to 1.4\unit{dB} for $m=4$.
This performance increase for higher constellation orders is also expected, as higher constellations require better estimation to ensure good symbol and bit decisions.

\section{Conclusion and Outlook}

We have shown that \ac{NN}-based estimation for time- and frequency-selective channels exhibits a robust behavior w.r.t. inaccurate channel parameter knowledge.
It turned out that \ac{NN}-based channel estimation can compete with the (close to) optimal \ac{LMMSE} estimator for a wide range of channel parameters and can even outperform the baseline for scenarios with inaccurate parameter estimations. 
When compared to the baseline without perfect velocity estimation and evaluated on a more realistic channel with \ac{LoS} components, we observed significant gains.
We have kept full compatibility to the IEEE 802.11p piloting scheme, however, extensions to other piloting schemes are straightforward. 
Finally, we have showcased the possibility of \emph{data-aware} equalization that is able to extract further statistical information of the channel from the payload data sequence leading to significant \ac{MSE} and consequently \ac{BER} performance gains.

It remains open for future work to evaluate (and potentially train) the system with real-world data from larger measurement campaigns, in particular, from V2X scenarios.
We want to emphasize that the main contribution of this work is not necessarily found in plain performance improvements for fixed channel setups, but to showcase the conceptual simplicity of such a data-driven and \ac{RNN}-based system design.

\bibliographystyle{IEEEtran}
\bibliography{IEEEabrv,references}

\begin{thebibliography}{10}
\providecommand{\url}[1]{#1}
\csname url@samestyle\endcsname
\providecommand{\newblock}{\relax}
\providecommand{\bibinfo}[2]{#2}
\providecommand{\BIBentrySTDinterwordspacing}{\spaceskip=0pt\relax}
\providecommand{\BIBentryALTinterwordstretchfactor}{4}
\providecommand{\BIBentryALTinterwordspacing}{\spaceskip=\fontdimen2\font plus
\BIBentryALTinterwordstretchfactor\fontdimen3\font minus
  \fontdimen4\font\relax}
\providecommand{\BIBforeignlanguage}[2]{{%
\expandafter\ifx\csname l@#1\endcsname\relax
\typeout{** WARNING: IEEEtran.bst: No hyphenation pattern has been}%
\typeout{** loaded for the language `#1'. Using the pattern for}%
\typeout{** the default language instead.}%
\else
\language=\csname l@#1\endcsname
\fi
#2}}
\providecommand{\BIBdecl}{\relax}
\BIBdecl

\bibitem{nilsson1997analysis}
R.~Nilsson, O.~Edfors, M.~Sandell, and P.~O. Borjesson, ``An analysis of
  two-dimensional pilot-symbol assisted modulation for {O}{F}{D}{M},'' in
  \emph{1997 IEEE International Conference on Personal Wireless Communications
  (Cat. No. 97TH8338)}.\hskip 1em plus 0.5em minus 0.4em\relax IEEE, 1997, pp.
  71--74.

\bibitem{hoeher1997two}
P.~Hoeher, S.~Kaiser, and P.~Robertson, ``Two-dimensional pilot-symbol-aided
  channel estimation by {W}iener filtering,'' in \emph{1997 IEEE International
  Conference on Acoustics, Speech, and Signal Processing}, vol.~3.\hskip 1em
  plus 0.5em minus 0.4em\relax IEEE, 1997, pp. 1845--1848.

\bibitem{dong2007linear}
X.~Dong, W.-S. Lu, and A.~C. Soong, ``Linear interpolation in pilot symbol
  assisted channel estimation for {O}{F}{D}{M},'' \emph{IEEE Transactions on
  Wireless Communications}, vol.~6, no.~5, pp. 1910--1920, 2007.

\bibitem{8054694}
T.~O'Shea and J.~Hoydis, ``{An Introduction to Deep Learning for the Physical
  Layer},'' \emph{IEEE Trans. Cogn. Commun. Netw.}, vol.~3, no.~4, pp.
  563--575, Dec. 2017.

\bibitem{farsad2018neural}
N.~Farsad and A.~Goldsmith, ``Neural network detection of data sequences in
  communication systems,'' \emph{IEEE Trans. on Signal Process.}, vol.~66,
  no.~21, pp. 5663--5678, 2018.

\bibitem{nachmani2016learning}
E.~Nachmani, Y.~Be'ery, and D.~Burshtein, ``Learning to decode linear codes
  using deep learning,'' in \emph{Allerton Conf.}\hskip 1em plus 0.5em minus
  0.4em\relax IEEE, 2016, pp. 341--346.

\bibitem{neumann2018learning}
D.~Neumann, T.~Wiese, and W.~Utschick, ``Learning the {M}{M}{S}{E} channel
  estimator,'' \emph{IEEE Transactions on Signal Processing}, vol.~66, no.~11,
  pp. 2905--2917, 2018.

\bibitem{luo2018channel}
C.~Luo, J.~Ji, Q.~Wang, X.~Chen, and P.~Li, ``Channel state information
  prediction for 5{G} wireless communications: A deep learning approach,''
  \emph{IEEE Transactions on Network Science and Engineering}, 2018.

\bibitem{mashhadi2020pruning}
M.~B. Mashhadi and D.~Gunduz, ``Pruning the pilots: Deep learning-based pilot
  design and channel estimation for {M}{I}{M}{O}-{O}{F}{D}{M} systems,''
  \emph{arXiv preprint arXiv:2006.11796}, 2020.

\bibitem{honkala2020deeprx}
M.~Honkala, D.~Korpi, and J.~M. Huttunen, ``Deeprx: Fully convolutional deep
  learning receiver,'' \emph{arXiv preprint arXiv:2005.01494}, 2020.

\bibitem{aoudia2020end}
F.~A. Aoudia and J.~Hoydis, ``End-to-end learning for {O}{F}{D}{M}: From neural
  receivers to pilotless communication,'' \emph{arXiv preprint
  arXiv:2009.05261}, 2020.

\bibitem{li2018power}
{H. Ye, G. Ye Li, B. Juang}, ``{Power of Deep Learning for Channel Estimation
  and Signal Detection in OFDM Systems},'' \emph{IEEE Wireless Communications
  Letters}, 2018.

\bibitem{tandler2019recurrent}
D.~Tandler, S.~D{\"o}rner, S.~Cammerer, and S.~ten Brink, ``On recurrent neural
  networks for sequence-based processing in communications,'' in \emph{2019
  53rd Asilomar Conference on Signals, Systems, and Computers}.\hskip 1em plus
  0.5em minus 0.4em\relax IEEE, 2019, pp. 537--543.

\bibitem{ieee2010802}
I.~S. Association \emph{et~al.}, ``802.11p-2010-{I}{E}{E}{E} standard for
  information technology-local and metropolitan area networks-specific
  requirements-part 11: Wireless {L}{A}{N} medium access control ({M}{A}{C})
  and physical layer ({P}{H}{Y}) specifications amendment 6: Wireless access in
  vehicular environments,'' 2010.

\bibitem{Molisch}
A.~F. Molisch, \emph{{Wireless Communications}}, 2nd~ed.\hskip 1em plus 0.5em
  minus 0.4em\relax John Wiley \& Sons Ltd., 2016.

\bibitem{Stueber}
G.~L. St{\"u}ber, \emph{{Principles of Mobile Communication}}, 4th~ed.\hskip
  1em plus 0.5em minus 0.4em\relax Springer International Publishing, 2017.

\bibitem{Xiao}
C.~{Xiao}, Y.~R. {Zheng}, and N.~C. {Beaulieu}, ``Statistical simulation models
  for {R}ayleigh and {R}ician fading,'' in \emph{IEEE International Conference
  on Communications, 2003. ICC '03.}, vol.~5, 2003, pp. 3524--3529 vol.5.

\bibitem{Simko_Kronecker}
M.~{Šimko}, C.~{Mehlführer}, M.~{Wrulich}, and M.~{Rupp}, ``Doubly dispersive
  channel estimation with scalable complexity,'' in \emph{2010 International
  ITG Workshop on Smart Antennas (WSA)}, 2010, pp. 251--256.

\bibitem{Hochreiter}
S.~Hochreiter and J.~Schmidhuber, ``Long short-term memory,'' \emph{Neural
  computation}, vol.~9, pp. 1735--80, 12 1997.

\bibitem{cammerer2020trainable}
S.~Cammerer, F.~A. Aoudia, S.~D{\"o}rner, M.~Stark, J.~Hoydis, and S.~ten
  Brink, ``Trainable communication systems: Concepts and prototype,''
  \emph{IEEE Transactions on Communications}, vol.~68, no.~9, pp. 5489--5503,
  2020.

\end{thebibliography}

\end{document}